\documentclass[titlepage,12pt]{article}
\usepackage{amssymb}
\usepackage{amsfonts}
\usepackage{geometry}
\usepackage{mathrsfs}
\usepackage[onehalfspacing]{setspace}

\newtheorem{theorem}{Theorem}

\input{tcilatex}
\renewcommand{\mathcal}{\mathscr}

\begin{document}

\title{Joint models for grid point and response processes in longitudinal
and functional data}
\author{Daniel Gervini and Tyler J Baur \\
Department of Mathematical Sciences\\
University of Wisconsin--Milwaukee}
\maketitle

\begin{abstract}
The distribution of the grid points at which a response function is observed
in longitudinal or functional data applications is often informative and not
independent of the response process. In this paper we introduce a
covariation model to estimate and make inferences about this interrelation,
by treating the data as replicated realizations of a marked point process.
We derive maximum likelihood estimators, the asymptotic distribution of the
estimators, and study the estimators' behavior by simulation. We apply the
model to an online auction data set and show that there is a strong
correlation between bidding patterns and price trajectories.

\end{abstract}

\section{Introduction\label{sec:Introd}}

In many statistical applications the object of analysis are samples of
functions, $\{g_{i}(x):i=1,\ldots ,n\}$. These functions are generally
measured at discrete points $\{x_{ij}:j=1,\ldots ,m_{i}\}$, so the data
actually observed is $\{(x_{ij},y_{ij}):j=1,\ldots ,m_{i},\ i=1,\ldots ,n\}$
with 
\begin{equation}
y_{ij}=g_{i}(x_{ij})+\eta _{ij},  \label{eq:model_y}
\end{equation}%
where $\eta _{ij}$ is random noise. Longitudinal data often fits this
framework (Rice, 2004; M\"{u}ller, 2008).

Functional data analysis has focused on the analysis of the functions $%
g_{i}(x)$s, which are usually recovered from the raw data by some form of
smoothing (James et al., 2000; Ramsay and Silverman, 2005, ch.~3; Yao et
al., 2005). The distribution of the grid points $\{x_{ij}\}$ is generally
considered noninformative. However, there are situations where the
distribution of the $x_{ij}$s may be informative in its own right.

Consider, for example, the bid price trajectories shown in Figure \ref%
{fig:samp_data}. They are bid prices of Palm M515 Personal Digital
Assistants (PDA) on week-long eBay auctions that took place between March
and May of 2003. Bidding activity tends to concentrate at the beginning and
at the end of the auctions, in patterns that have been called `early
bidding'\ and `bid sniping', respectively. Earlier analyses of these data
(Shmueli and Jank, 2005; Jank and Shmueli, 2006, 2010) studied the dynamics
of the process via derivatives of the bid price curves. More recently, Wu et
al.~(2013) and Arribas-Gil and M\"{u}ller (2014) investigated the bid time
process itself. But a joint modeling of the bid time process and the bid
price curves has not been attempted, and there are reasons to believe these
processes are not independent. For example, it is suspected that items with
prices below the mean are more likely to experience `bid sniping'. To study
such questions it is necessary to jointly model the bid time process $%
\{x_{ij}\}$ and the bid price process $\{y_{ij}\}$.

The approach we present in this paper considers the data $%
\{(x_{ij},y_{ij})\} $ as $n$ independent realizations of a marked point
process. For each subject $i$, the $x_{ij}$s are seen as a realization of a
point process and the $y_{ij}$s as the corresponding `marks', to use common
point-process terminology (Cox and Isham, 1980; M\o ller and Waagepetersen,
2004; Baddeley, 2007; Streit, 2010). Note, however, that not all marked
point processes arise as discretizations of smooth functions as in model (%
\ref{eq:model_y}); the methods we propose here are specifically intended for
functional and longitudinal data applications. To avoid confusions with
terminology, we will not call the $m_{i}$ observations for each subject $i$
`replications', as is often done in the point process literature; we
consider the whole set $\{(x_{ij},y_{ij}):j=1,\ldots ,m_{i}\}$ for each $i$
as a single realization of the process, and the $n$ replications are the
sets for $i=1,\ldots ,n$.

As pointed out by Guan and Afshartous (2007) and M\o ller et al.~(2016), the
literature on modeling marked point processes is limited, and restricted to
the single replication scenario; it has focused on simple summary statistics
of the processes and on testing broad generic hypotheses such as independent
marking (Guan and Afshartous, 2007; Myllym\"{a}ki et al., 2017; see also
Baddeley, 2010, sec.~21.7). But the availability of replications allows us
to estimate the correlations between the intensity functions of the point
process $\{x_{ij}\}$ and the Karhunen--Lo\`{e}ve components of the response
process $\{y_{ij}\}$, which is not possible in a single-replication
scenario. Regression models in point process contexts have been proposed
recently (Barret et al., 2015; Rathbun and Shiffman, 2016), but they aim at
incorporating covariates into intensity function models. Similarly, Scheike
(1997) related longitudinal data to marked point processes, but his goal was
to model the conditional distribution of the time points given the past
observations. None of those papers aim at jointly modeling the time points
and the response processes, which is the goal of this paper.

\section{Latent variable model\label{sec:Model}}

A point process $X$ is a random countable set in a space $\mathcal{S}$,
where $\mathcal{S}$ is usually $\mathbb{R}$ for temporal processes or $%
\mathbb{R}^{2}$ for spatial processes (M\o ller and Waagepetersen, 2004,
ch.~2; Streit, 2010, ch.~2). When each point $x\in X$ is accompanied by a
random feature $Y_{x}$ in some space $\mathcal{M}$, $Z=\{(x,Y_{x}):x\in X\}$
is called a marked point process. As mentioned in Section \ref{sec:Introd},
we are interested in the specific situation where $Y_{x}$ follows the model 
\begin{equation}
Y_{x}=g(x)+\eta _{x},  \label{eq:model_Yx}
\end{equation}%
with $g:\mathcal{S}\rightarrow \mathcal{M}$ the function of interest and $%
\eta _{x}$ random noise. We will consider only $\mathcal{M}=\mathbb{R}$ in
this paper, but extensions to the multivariate case $\mathcal{M}=\mathbb{R}%
^{k}$ are straightforward.

A point process $X$ is locally finite if $\#(X\cap B)<\infty $ with
probability one for any bounded $B\subseteq \mathcal{S}$. For a locally
finite process the count function $N(B)=\#(X\cap B)$ can be defined, and $%
Z_{B}:=\{(x,Y_{x}):x\in X\cap B\}$ is a finite set, $Z_{B}=\{(x_{1},y_{1}),%
\ldots ,(x_{N(B)},y_{N(B)})\}$. A Poisson process is a locally finite
process for which there exists a locally integrable function $\lambda :%
\mathcal{S}\rightarrow \lbrack 0,\infty )$, called the intensity function,
such that (i) $N(B)$ has a Poisson distribution with rate $\int_{B}\lambda
(t)dt$, and (ii) for disjoint sets $B_{1},\ldots ,B_{k}$ the random
variables $N(B_{1}),\ldots ,N(B_{k})$ are independent. A consequence of (i)
and (ii) is that the conditional distribution of the points in $X\cap B$
given $N(B)=m$ is the distribution of $m$ independent and identically
distributed observations with density $\lambda (t)/\int_{B}\lambda $.

For replicated point processes, a single intensity function $\lambda $
rarely provides an adequate fit for all replications. It is more reasonable
to assume that the $\lambda $s are subject-specific and treat them as random
effects. Such processes are called doubly stochastic processes or Cox
processes (M\o ller and Waagepetersen, 2004, ch.~5; Streit, 2010, ch.~8). A
doubly stochastic process is a pair $(X,\Lambda )$ where $X|\Lambda =\lambda 
$ is a Poisson process with intensity function $\lambda $, and $\Lambda $ is
a random function that takes values on the space $\mathcal{F}$ of
non-negative locally integrable functions on $\mathcal{S}$. Then the $n$
replications of the point process can be seen as independent identically
distributed~realizations of a doubly stochastic process $(X,\Lambda )$,
where $X$ is observable but $\Lambda $ is not. Similarly, for $g$ in (\ref%
{eq:model_Yx}) we will assume there is a process $G$ such that $Y\mid
(X,G=g) $ follows model (\ref{eq:model_Yx}). Then the $n$ replications of
the marked point process can be seen as independent identically
distributed~realizations of $(X,Y,\Lambda ,G)$, where $X$ and $Y$ are
observable but $\Lambda $ and $G$ are not.

Our main goal is to study the relationship between the intensity process $%
\Lambda $ that generates the $x$s and the response process $G$ that
generates the $y$s. To this end we will assume that $G$ follows a finite
Karhunen--Lo\`{e}ve decomposition 
\begin{equation}
G(x)=\nu (x)+\sum_{k=1}^{p_{2}}v_{k}\psi _{k}(x),  \label{eq:PC_G}
\end{equation}%
where the $\psi _{k}$s are orthonormal functions in $L^{2}(\mathcal{S})$ and
the $v_{k}$s are uncorrelated zero-mean random variables. Any stochastic
process in $L^{2}(\mathcal{S})$ with finite variance can be decomposed as in
(\ref{eq:PC_G}) with a possibly infinite $p_{2}$ (Ash and Gardner, 1975,
ch.~1.4), but since we are interested in smooth processes in this paper, for
practical purposes it is sufficient to consider only finite $p_{2}$s.

A similar decomposition for $\Lambda $ would be problematic due to the
non-negativity constraint. A nonnegative decomposition was proposed by
Gervini (2016). However, for simplicity we will use an alternative approach
in this paper, and decompose instead the logarithm of $\Lambda $, which is
unconstrained: 
\begin{equation}
\log \Lambda (x)=\mu (x)+\sum_{k=1}^{p_{1}}u_{k}\phi _{k}(x),
\label{eq:Lambda_exp}
\end{equation}%
where the $\phi _{k}$s are orthonormal functions in $L^{2}(\mathcal{S})$ and
the $u_{k}$s are uncorrelated zero-mean random variables.

The association between $\Lambda $ and $G$ is then determined by the
association between the component scores $\mathbf{u}=(u_{1},\ldots
,u_{p_{1}})$ and $\mathbf{v}=(v_{1},\ldots ,v_{p_{2}})$ in (\ref{eq:PC_G})
and (\ref{eq:Lambda_exp}). As a working model, we will assume that $(\mathbf{%
u},\mathbf{v})$ follows a joint multivariate normal distribution with mean
zero and covariance matrix 
\[
\mathbf{\Sigma }=\left( 
\begin{array}{cc}
\func{diag}(\mathbf{\sigma }_{u}^{2}) & \mathbf{\Sigma }_{uv} \\ 
\mathbf{\Sigma }_{uv}^{T} & \func{diag}(\mathbf{\sigma }_{v}^{2})%
\end{array}%
\right) , 
\]%
where $\mathbf{\sigma }_{u}^{2}$ and $\mathbf{\sigma }_{v}^{2}$ are the
variances of the $u_{k}$s and the $v_{k}$s, respectively. The error term $%
\eta $ in (\ref{eq:model_Yx}) is assumed $N(0,\sigma _{\eta }^{2})$ and
independent of the $u_{k}$s and the $v_{k}$s. The parameter of interest here
is the cross-covariance matrix $\mathbf{\Sigma }_{uv}$; the others are
mostly nuisance parameters.

The signs of the component scores are not identifiable, since $-u_{k}$ and $%
-\phi _{k}(x)$ satisfy the same model as $u_{k}$ and $\phi _{k}(x)$, and
similarly with the $v_{k}$s and $\psi _{k}$s. Consequently, the signs of $%
\Sigma _{uv,kl}=\limfunc{cov}(u_{k},v_{l})$ are not identifiable either and
can be chosen for convenience of interpretation for any given application.

To facilitate estimation of the functional parameters $\mu $, $\phi _{k}$s, $%
\nu $ and $\psi _{k}$s, we will use semiparametric basis-function
expansions. As basis functions one can take, for instance, B-splines if $%
\mathcal{S}=\mathbb{R}$, or normalized Gaussian radial kernels if $\mathcal{S%
}=\mathbb{R}^{2}$; other families are possible and perhaps better in some
cases, such as simplicial bases for bivariate functions on irregular
domains. We will call this family $\mathcal{B}$. Let $\mathbf{\gamma }(x)$
be the vector of basis functions $\{\gamma _{1},\ldots ,\gamma _{q}\}$ of $%
\mathcal{B}$, with $\gamma _{j}:\mathcal{S}\rightarrow \mathbb{R}$. We
assume, then, that $\mu (x)=\mathbf{c}_{0}^{T}\mathbf{\gamma }(x)$, $\phi
_{k}(x)=\mathbf{c}_{k}^{T}\mathbf{\gamma }(x)$, $\nu (x)=\mathbf{d}_{0}^{T}%
\mathbf{\gamma }(x)$ and $\psi _{k}(x)=\mathbf{d}_{k}^{T}\mathbf{\gamma }(x)$%
.

The model parameters will be collected, for simplicity, in a single vector 
\begin{equation}
\mathbf{\theta }=(\func{vec}\mathbf{\Sigma }_{uv},\mathbf{c}_{0},\ldots ,%
\mathbf{c}_{p_{1}},\mathbf{d}_{0},\ldots ,\mathbf{d}_{p_{2}},\mathbf{\sigma }%
_{u}^{2},\mathbf{\sigma }_{v}^{2},\sigma _{\eta }^{2}).  \label{eq:theta}
\end{equation}%
The orthonormality constraints on the $\phi _{k}$s and the $\psi _{k}$s can
be expressed as $\mathbf{c}_{k}^{T}\mathbf{Jc}_{l}=\mathbf{d}_{k}^{T}\mathbf{%
Jd}_{l}=\delta _{kl}$, where $\delta _{kl}$ is Kronecker's delta and $%
\mathbf{J}=\int \mathbf{\gamma }(x)\mathbf{\gamma }(x)^{T}dx$.

\section{Penalized maximum likelihood estimation\label{sec:Estimation}}

With a slight abuse of notation, let us write $\{(x_{ij},y_{ij}):j=1,\ldots
,m_{i}\}$ in vector form, $(\mathbf{x}_{i},m_{i},\mathbf{y}_{i})$. Then the
joint density of observations and latent variables can be factorized as 
\[
f_{\mathbf{\theta }}(\mathbf{x},m,\mathbf{y},\mathbf{u},\mathbf{v})=f_{%
\mathbf{\theta }}(\mathbf{y}\mid \mathbf{x},m,\mathbf{u},\mathbf{v})f_{%
\mathbf{\theta }}(\mathbf{x},m\mid \mathbf{u},\mathbf{v})f_{\mathbf{\theta }%
}(\mathbf{u},\mathbf{v}). 
\]%
Since $f_{\mathbf{\theta }}(\mathbf{y}\mid \mathbf{x},m,\mathbf{u},\mathbf{v}%
)$ does not explicitly depend on $\mathbf{u}$ and $f_{\mathbf{\theta }}(%
\mathbf{x},m\mid \mathbf{u},\mathbf{v})$ does not explicitly depend on $%
\mathbf{v}$, we can write 
\[
f_{\mathbf{\theta }}(\mathbf{x},m,\mathbf{y},\mathbf{u},\mathbf{v})=f_{%
\mathbf{\theta }}(\mathbf{y}\mid \mathbf{x},m,\mathbf{v})f_{\mathbf{\theta }%
}(\mathbf{x},m\mid \mathbf{u})f_{\mathbf{\theta }}(\mathbf{u},\mathbf{v}). 
\]%
From (\ref{eq:model_Yx}), (\ref{eq:PC_G}), (\ref{eq:Lambda_exp}) and the
distributional assumptions made in Section \ref{sec:Model}, we have: 
\begin{equation}
f_{\mathbf{\theta }}(\mathbf{y}\mid \mathbf{x},m,\mathbf{v})=\frac{1}{(2\pi
\sigma _{\eta }^{2})^{m/2}}\exp \left\{ -\frac{1}{2\sigma _{\eta }^{2}}\Vert 
\mathbf{y}-\nu (\mathbf{x})-\mathbf{\Psi }(\mathbf{x})\mathbf{v}\Vert
^{2}\right\} ,  \label{eq:Cond_fy}
\end{equation}%
with $\nu (\mathbf{x})=(\nu (x_{1}),\ldots ,\nu (x_{m}))^{T}$ and $\mathbf{%
\Psi }(\mathbf{x})=[\psi _{1}(\mathbf{x}),\ldots ,\psi _{p_{2}}(\mathbf{x})]$%
; 
\[
f_{\mathbf{\theta }}(\mathbf{x},m\mid \mathbf{u})=\exp \left\{ -\int \lambda
_{\mathbf{u}}(t)dt\right\} \frac{1}{m!}\prod_{j=1}^{m}\lambda _{\mathbf{u}%
}(x_{j}), 
\]%
with $\lambda _{\mathbf{u}}(x)=\exp \{\mu (x)+\mathbf{u}^{T}\mathbf{\phi }%
(x)\}$; and 
\[
f_{\mathbf{\theta }}(\mathbf{u},\mathbf{v})=\frac{1}{(2\pi
)^{(p_{1}+p_{2})/2}(\det \mathbf{\Sigma })^{1/2}}\exp \left\{ -\frac{1}{2}(%
\mathbf{u}^{T},\mathbf{v}^{T})\mathbf{\Sigma }^{-1}(\mathbf{u}^{T},\mathbf{v}%
^{T})^{T}\right\} . 
\]%
The marginal density of the observations, 
\[
f_{\mathbf{\theta }}(\mathbf{x},m,\mathbf{y})=\iint f_{\mathbf{\theta }}(%
\mathbf{x},m,\mathbf{y},\mathbf{u},\mathbf{v})~d\mathbf{u}~d\mathbf{v,} 
\]%
has no closed form and requires numerical integration for its evaluation,
for which we use the Laplace approximation. This and other details of
implementation are discussed in the Supplementary Material.

The maximum likelihood estimator of $\mathbf{\theta }$ would be the
maximizer of $\sum_{i=1}^{n}\log f_{\mathbf{\theta }}(\mathbf{x}_{i},m_{i},%
\mathbf{y}_{i})$. However, when a large family of basis functions $\mathcal{B%
}$ is used, it is advisable to regularize the functional estimators by
adding roughness penalties to the objective function. So we define the
penalized log-likelihood 
\begin{equation}
\ell _{n}(\mathbf{\theta })=\frac{1}{n}\sum_{i=1}^{n}\log f_{\mathbf{\theta }%
}(\mathbf{x}_{i},m_{i},\mathbf{y}_{i})-\xi _{1}P(\mu )-\xi
_{2}\sum_{k=1}^{p_{1}}P(\phi _{k})-\xi _{3}P(\nu )-\xi
_{4}\sum_{k=1}^{p_{2}}P(\psi _{k}),  \label{eq:Pen-log-lik}
\end{equation}%
where the $\xi $s are nonnegative smoothing parameters and $P(f)$ is a
roughness penalty function, such as $P(f)=\int (f^{\prime \prime })^{2}$ if $%
f$ is univariate or $P(f)=\iint \{(\frac{\partial ^{2}f}{\partial t_{1}^{2}}%
)^{2}+2(\frac{\partial ^{2}f}{\partial t_{1}\partial t_{2}})^{2}+(\frac{%
\partial ^{2}f}{\partial t_{2}^{2}})^{2}\}$ if $f$ is bivariate. The
estimator of $\mathbf{\theta }$ is then defined as 
\[
\mathbf{\hat{\theta}}=\arg \max_{\mathbf{\theta }\in \Theta }\ell _{n}(%
\mathbf{\theta }), 
\]%
where $\Theta $ is the parameter space 
\begin{eqnarray}
\Theta &=&\{\mathbf{\theta }\in \mathbb{R}^{d}:h_{kl}^{C}(\mathbf{\theta }%
)=0,\ \ k=1,\ldots ,l,\ \ l=1,\ldots ,p_{1},  \label{eq:Theta_prelim} \\
&&h_{kl}^{D}(\mathbf{\theta })=0,\ \ k=1,\ldots ,l,\ \ l=1,\ldots ,p_{2}, 
\nonumber \\
&&\sigma _{\eta }^{2}>0,\ \mathbf{\Sigma }>0\},  \nonumber
\end{eqnarray}%
with $d$ the dimension of $\mathbf{\theta }$, $h_{kl}^{C}(\mathbf{\theta })=%
\mathbf{c}_{k}^{T}\mathbf{Jc}_{l}-\delta _{kl}$, $h_{kl}^{D}(\mathbf{\theta }%
)=\mathbf{d}_{k}^{T}\mathbf{Jd}_{l}-\delta _{kl}$, and $\mathbf{\Sigma }>0$
denoting that $\mathbf{\Sigma }$ is symmetric and positive definite. The
estimating equations for $\mathbf{\hat{\theta}}$ and an EM algorithm
(Dempster et al., 1977) for its computation are derived in the Supplementary
Material. The programs implementing these algorithms are available on the
first author's website.

Once $\mathbf{\hat{\theta}}$ has been obtained, individual predictors of the
latent component scores, whether for the sample units or for new data, can
be obtained as $\mathbf{\hat{u}}_{i}=E_{\mathbf{\hat{\theta}}}(\mathbf{u}%
\mid \mathbf{x}_{i},m_{i},\mathbf{y}_{i})$ and $\mathbf{\hat{v}}_{i}=E_{%
\mathbf{\hat{\theta}}}(\mathbf{v}\mid \mathbf{x}_{i},m_{i},\mathbf{y}_{i})$.
These integrals can also be numerically evaluated by Laplace approximation.

This model has a number of tuning parameters that have to be chosen by the
user: the number of functional components $p_{1}$ and $p_{1}$, the type of
basis family $\mathcal{B}$ and its dimension $q$, and the smoothing
parameters $\xi $s in the penalized likelihood. The specific type of basis
family will not have much of an impact for most applications, provided the
dimension $q$ is large enough. In this paper we use cubic $B$-splines with
equally spaced knots for our simulations and data analyses; higher-order
splines should be used if estimation of derivatives is of interest. The
dimension $q$ is more relevant and should be relatively large to avoid bias;
the variability of the estimators will be taken care of by the $\xi $s. As
noted by Ruppert (2002, sec.~3), although $q$ can be chosen systematically
by cross-validation, there is little change in goodness of fit after a
minimum dimension $q$ has been reached, and the fit will essentially be
determined by the smoothing parameters thereafter.

The choice of $\xi $s, then, is more important, and can be done objectively
by cross-validation (Hastie \emph{et al.}, 2009, ch.~7). Leave-one-out
cross-validation finds $\xi $s that maximize 
\begin{equation}
\mathrm{CV}(\xi _{1},\xi _{2},\xi _{3},\xi _{4})=\sum_{i=1}^{n}\log f_{%
\mathbf{\hat{\theta}}^{[-i]}}(\mathbf{x}_{i},m_{i},\mathbf{y}_{i}),
\label{eq:CV}
\end{equation}%
where $\mathbf{\hat{\theta}}^{[-i]}$ denotes the estimator obtained without
observation $i$. A faster alternative is to use $k$-fold cross-validation,
where the data is split into $k$ subsets that are alternatively used as test
data; $k=5$ is a common choice. Full four-dimensional optimization of (\ref%
{eq:CV}) would be too time consuming even with five-fold cross-validation,
so as a workable alternative we suggest a sequential optimization, where
each $\xi _{j}$ in turn is optimized on a grid while the others are kept
fixed at an initial value chosen by the user.

A more practical alternative is to choose the parameters subjectively by
visual inspection. Plots of the means and components for different $\xi $s
on a grid can be inspected to see how new features of the curves appear or
disappear as $\xi $ varies, and choose $\xi $s that produce curves with
well-defined but not too irregular features. In general, since curve shapes
change smoothly with $\xi $, there is a relatively broad range of $\xi $s
that will produce reasonable results; it is not necessary to specify a
precise optimal. We use this method in our simulations and data analysis in
this paper.

The choice of the number of components $p_{1}$ and $p_{2}$ can also be done
either objectively by cross-validation or subjectively by taking into
account the accumulated proportions of variability $\sigma _{u1}^{2}+\cdots
+\sigma _{up_{1}}^{2}$ and $\sigma _{v1}^{2}+\cdots +\sigma _{vp_{2}}^{2}$.
From a practical perspective, however, the goal of this model is not so much
to find the largest possible $p$s that will best approximate the data, but
to capture the most salient modes of variability of the $X$ and $Y$
processes and estimate and interpret their correlations; from this
perspective, a few well-estimated components with significant correlations
will be preferable to a higher-dimensional model without many (or any)
significant correlations, even if some residual systematic variability
remains unaccounted for.

\section{Asymptotics and inference\label{sec:Asymptotics}}

The asymptotic behavior of $\mathbf{\hat{\theta}}$ as $n\rightarrow \infty $
can be studied via standard empirical-process techniques (Pollard, 1984; Van
der Vaart, 2000), since (\ref{eq:Pen-log-lik}) is the average of independent
identically distributed functions plus a non-random roughness penalty, as in
e.g.~Knight and Fu (2000).

A `nonparametric'\ asymptotics where no assumptions about the functional
parameters (other than degrees of smoothness) are made and the dimension $q$
of the basis family $\mathcal{B}$ is allowed to grow with $n$ is perhaps the
most theoretically satisfying, but it is too difficult. A simpler approach
is the `parametric' asymptotics, where $q$ is held fixed and the functional
parameters are assumed to belong to $\mathcal{B}$. This approach, in effect,
ignores smoothing bias, but in practice this is not a serious problem as
long as $q$ is reasonably large. We will then follow this approach, which
others have followed in similar semiparametric contexts (e.g.~Yu and
Ruppert, 2002, and Xun et al., 2013), and show later by simulation that the
asymptotic variance estimates provide very accurate approximations to the
actual finite-sample variance of the estimators.

The first result in this section, Theorem \ref{thm:Constcy}, establishes
consistency of the estimator $\mathbf{\hat{\theta}}$. The proof, given in
the Supplementary Material, essentially follows along the lines of the
classical consistency proof of maximum likelihood estimators, with the
caveat that the indeterminate sign of the functional components requires
special handling. We will also assume that the components have multiplicity
one, so we define 
\begin{eqnarray}
\Theta &=&\{\mathbf{\theta }\in \mathbb{R}^{s}:h_{kl}^{C}(\mathbf{\theta }%
)=0,\ \ k=1,\ldots ,l,\ \ l=1,\ldots ,p_{1},  \label{eq:Theta_final} \\
&&h_{kl}^{D}(\mathbf{\theta })=0,\ \ k=1,\ldots ,l,\ \ l=1,\ldots ,p_{2}, 
\nonumber \\
&&\sigma _{\eta }^{2}>0,\ \mathbf{\Sigma }>0,\ \sigma _{u1}>\cdots >\sigma
_{up_{1}}>0,\ \sigma _{v1}>\cdots >\sigma _{vp_{2}}>0,  \nonumber \\
&& c_{k1} \geq 0,\ k=1,\ldots ,p_{1},\ d_{k1}\geq 0,\ k=1,\ldots ,p_{2}\}, 
\nonumber
\end{eqnarray}%
and make the following assumptions:

\begin{description}
\item[A1] The signs of the functional components $\hat{\phi}_{k,n}$ and $%
\hat{\psi}_{k,n}$ are specified so that the first non-zero basis coefficient
of each $\hat{\phi}_{k,n}$ and $\hat{\psi}_{k,n}$ is positive (then $\mathbf{%
\hat{\theta}}_{n}\in \Theta $ for $\Theta $ defined in (\ref{eq:Theta_final}%
).)

\item[A2] The true functional parameters $\mu _{0}$, $\nu _{0}$, $\phi _{k0}$%
s and $\psi _{k0}$s of model (\ref{eq:model_y})-(\ref{eq:PC_G})-(\ref%
{eq:Lambda_exp}) belong to the functional space $\mathcal{B}$ used for
estimation, and the basis coefficients $c_{k1,0}$ and $d_{k1,0}$ are not
zero. The signs of $\phi _{k0}$ and $\psi _{k0}$ are then specified so that $%
c_{k1,0}>0$ and $d_{k1,0}>0$; therefore there is a unique $\mathbf{\theta }%
_{0}$ in $\Theta $ such that $f_{\mathbf{\theta }_{0}}(\mathbf{x},m,\mathbf{y%
})$ is the true density of the data.

\item[A3] $\mathbf{\xi }_{n}\rightarrow \mathbf{0}$ as $n\rightarrow \infty $%
, where $\mathbf{\xi }_{n}=(\xi _{1n},\xi _{2n},\xi _{3n},\xi _{4n})^{T}$ is
the vector of smoothing parameters in (\ref{eq:Pen-log-lik}).
\end{description}

The requirement in assumption A2 that the first basis coefficients $c_{k1,0}$
and $d_{k1,0}$ of each $\phi _{k0}$ and $\psi _{k0}$ be non-zero, and
therefore can be taken strictly positive, is somewhat artificial; clearly
the $\phi _{k0}$s and $\psi _{k0}$s must have at least one non-zero basis
coefficient, but it need not be the first one or anyone else in particular.
However, some condition like this is necessary to uniquely identify a\
`true' parameter $\mathbf{\theta }_{0}$, which would otherwise be
unidentifiable due to sign ambiguity. This condition has to be consistent
with the sign-specification rule for the estimators in assumption A1.

\begin{theorem}
\label{thm:Constcy}Under assumptions A1--A3, $\mathbf{\hat{\theta}}_{n}%
\overset{P}{\rightarrow}%
\mathbf{\theta }_{0}$ as $n\rightarrow \infty $.
\end{theorem}

To establish the asymptotic normality of the estimators we follow the
approach of Geyer (1994), which makes use of the tangent cone of the
parameter space. The definition and properties of tangent cones can be found
in Rockafellar and Wets (1998, ch.~6). Using Theorem 6.31 of Rockafellar and
Wets (1998), the tangent cone of $\Theta $ at $\mathbf{\theta }_{0}$ is 
\begin{eqnarray*}
\mathcal{T}_{0} &=&\{\mathbf{\delta }\in \mathbb{R}^{s}:\nabla h_{kl}^{C}(%
\mathbf{\theta }_{0})^{T}\mathbf{\delta }=0,\ \ k=1,\ldots ,l,\ \ l=1,\ldots
,p_{1}, \\
&&\nabla h_{kl}^{D}(\mathbf{\theta }_{0})^{T}\mathbf{\delta }=0,\ \
k=1,\ldots ,l,\ \ l=1,\ldots ,p_{2}\}.
\end{eqnarray*}%
The explicit forms of $\nabla h_{kl}^{C}(\mathbf{\theta })$ and $\nabla
h_{kl}^{D}(\mathbf{\theta })$ are derived in the Supplementary Material. Let 
$\mathbf{A}$ be the $s_{1}\times s$ matrix with rows $\nabla h_{kl}^{C}(%
\mathbf{\theta }_{0})^{T}$ and $\nabla h_{kl}^{D}(\mathbf{\theta }_{0})^{T}$%
, where $s_{1}=\{p_{1}(p_{1}+1)/2+p_{2}(p_{2}+1)/2\}$, and let $\mathbf{B}$
be an orthogonal complement of $\mathbf{A}$, that is, an orthogonal $%
(s-s_{1})\times s$ matrix such that $\mathbf{AB}^{T}=\mathbf{O}$.

The next theorem gives the asymptotic distribution of $\mathbf{\hat{\theta}}%
_{n}$. In addition to $\mathbf{B}$ defined above, it uses Fisher's
information matrix, 
\begin{eqnarray*}
\mathbf{F}_{0} &=&E_{\mathbf{\theta }_{0}}\{\nabla \log f_{\mathbf{\theta }%
_{0}}(\mathbf{x},m,\mathbf{y})\nabla \log f_{\mathbf{\theta }_{0}}(\mathbf{x}%
,m,\mathbf{y})^{T}\} \\
&=&-E_{\mathbf{\theta }_{0}}\{\nabla ^{2}\log f_{\mathbf{\theta }_{0}}(%
\mathbf{x},m,\mathbf{y})\},
\end{eqnarray*}%
where $\nabla $ and $\nabla ^{2}$ are taken with respect to the parameter $%
\mathbf{\theta }$, and $\mathsf{D}\mathbf{P}(\mathbf{\theta })$, the
Jacobian matrix of the smoothness penalty vector $\mathbf{P}(\mathbf{\theta }%
)=(P(\mu ),\sum_{k=1}^{p_{1}}P(\phi _{k}),P(\nu ),\sum_{k=1}^{p_{2}}P(\psi
_{k}))^{T}$ of (\ref{eq:Pen-log-lik}). Explicit expressions for these
derivatives are given in the Supplementary Material. We make an additional
assumption:

\begin{description}
\item[A4] $\sqrt{n}\mathbf{\xi }_{n}\rightarrow \mathbf{\kappa }$ as $%
n\rightarrow \infty $, for a finite $\mathbf{\kappa }$.
\end{description}

\begin{theorem}
\label{thm:Asymp}Under assumptions A1--A4, $\sqrt{n}(\mathbf{\hat{\theta}}%
_{n}-\mathbf{\theta }_{0})%
\overset{D}{\rightarrow}%
\mathrm{N}(\mathbf{-V}\mathsf{D}\mathbf{P}(\mathbf{\theta }_{0})^{T}\mathbf{%
\kappa },\mathbf{V})$ as $n\rightarrow \infty $, with $\mathbf{V}=\mathbf{B}%
^{T}(\mathbf{BF}_{0}\mathbf{B}^{T})^{-1}\mathbf{B}$.
\end{theorem}

Fisher's information matrix $\mathbf{F}_{0}$ can be estimated by 
\[
\mathbf{\hat{F}}_{0}=\frac{1}{n}\sum_{i=1}^{n}\nabla \log f_{\mathbf{\hat{%
\theta}}}(\mathbf{x}_{i},m_{i},\mathbf{y}_{i})\nabla \log f_{\mathbf{\hat{%
\theta}}}(\mathbf{x}_{i},m_{i},\mathbf{y}_{i})^{T} 
\]%
and $\mathbf{V}$ by $\mathbf{\hat{V}}=\mathbf{B}^{T}(\mathbf{B\hat{F}}_{0}%
\mathbf{B}^{T})^{-1}\mathbf{B}$. The accuracy of the approximation of $%
\mathbf{\hat{V}}$ to the actual finite-sample variance of the estimators
depends on the ratio $n/s$. We found in our simulations (Section \ref%
{sec:Simulations}) that ratios $n/s\geq 3$ offer very accurate
approximations. This imposes some limitations on how large the basis family
dimension $q$ and the number of components $p_{1}$ and $p_{2}$ can be for
any given $n$.

\section{Simulations\label{sec:Simulations}}

We studied the finite sample behavior of the estimators by simulation,
assessing their consistency as the sample size increases and the goodness of
the approximation of the asymptotic variances.

We generated data from model (\ref{eq:model_Yx})-(\ref{eq:PC_G})-(\ref%
{eq:Lambda_exp}) with $p_{1}=p_{2}=2$. We considered a temporal process on $%
\mathcal{S}=[0,1]$, with $\mu (x)\equiv \sin \pi x-\log 1.98+\log r$, $\nu
(x)=5x$, $\phi _{1}(x)=\sqrt{2}\sin \pi x$, $\phi _{2}(x)=\sqrt{2}\sin 2\pi
x $, $\psi _{1}(x)=\phi _{1}(x)$ and $\psi _{2}(x)=\phi _{2}(x)$. The
baseline intensity function $\lambda _{0}(x)=\exp \mu (x)$ integrates to $r$%
; we chose two different values, $r=10$ and $r=30$, giving expected numbers
of observations per curve $10.5$ and $31.3$, respectively. The lower rate $%
r=10$ corresponds to the sparse situation where most individual trajectories
cannot be recovered by smoothing. The first components $\phi _{1}$ and $\psi
_{1}$ are essentially size components, explaining variation in overall level
above or below the mean, whereas the second components $\phi _{2}$ and $\psi
_{2}$ are contrasts, where e.g.~a positive score corresponds to curves that
are above the mean on the first half of $\mathcal{S}$ and below the mean on
the second half.

The component variances were of the form $\sigma _{u1}^{2}=.3^{2}\alpha $, $%
\sigma _{u2}^{2}=.3^{2}(1-\alpha )$, $\sigma _{v1}^{2}=.7^{2}\alpha $ and $%
\sigma _{v2}^{2}=.7^{2}(1-\alpha )$. Two choices of $\alpha $ were
considered: $\alpha =.60$ and $\alpha =.75$. The cross-covariance matrix $%
\mathbf{\Sigma }_{uv}$ was diagonal with elements $\mathbf{\Sigma }%
_{uv,11}=.7\sigma _{u1}\sigma _{v1}$ and $\mathbf{\Sigma }_{uv,22}=.7\sigma
_{u2}\sigma _{v2}$. The random-noise variance was $\sigma _{\eta
}^{2}=.3^{2} $. We considered four sample sizes $n$: $50$, $100$, $200$ and $%
400$. The combinations of $r$s, $\alpha $s and $n$s give us a total of 16
sampling models.

For estimation, we considered cubic $B$-spline families with five and ten
equally spaced knots. The smoothing parameters were visually chosen, as
explained in Section \ref{sec:Estimation}, from a few trial samples from
each of the six models with $r=10$ and each of the two knot sequences; the
same smoothing parameters were used for the respective models with $r=30$.
They are listed in the Supplementary Material. The Monte Carlo study, then,
considered a total of 32 scenarios, with two families of estimators per
sampling model. Each scenario was replicated 300 times.

\begin{table}[tbp] \centering%

\begin{tabular}{crcccrccc}
&  & \multicolumn{3}{c}{$r=10$} &  & \multicolumn{3}{c}{$r=30$} \\ 
Parameter & \multicolumn{1}{c}{} & $n=50$ & $n=100$ & $n=200$ & 
\multicolumn{1}{c}{} & $n=50$ & $n=100$ & $n=200$ \\ 
&  & \multicolumn{1}{r}{} & \multicolumn{1}{r}{} & \multicolumn{1}{r}{} &  & 
\multicolumn{1}{r}{} & \multicolumn{1}{r}{} & \multicolumn{1}{r}{} \\ 
$\Sigma _{uv,11}$ &  & \multicolumn{1}{r}{$.054$} & \multicolumn{1}{r}{$.031$%
} & \multicolumn{1}{r}{$.025$} &  & \multicolumn{1}{r}{$.038$} & 
\multicolumn{1}{r}{$.026$} & \multicolumn{1}{r}{$.019$} \\ 
$\Sigma _{uv,21}$ &  & \multicolumn{1}{r}{$.057$} & \multicolumn{1}{r}{$.038$%
} & \multicolumn{1}{r}{$.024$} &  & \multicolumn{1}{r}{$.028$} & 
\multicolumn{1}{r}{$.017$} & \multicolumn{1}{r}{$.011$} \\ 
$\Sigma _{uv,12}$ &  & \multicolumn{1}{r}{$.036$} & \multicolumn{1}{r}{$.023$%
} & \multicolumn{1}{r}{$.015$} &  & \multicolumn{1}{r}{$.021$} & 
\multicolumn{1}{r}{$.014$} & \multicolumn{1}{r}{$.010$} \\ 
$\Sigma _{uv,22}$ &  & \multicolumn{1}{r}{$.023$} & \multicolumn{1}{r}{$.017$%
} & \multicolumn{1}{r}{$.012$} &  & \multicolumn{1}{r}{$.014$} & 
\multicolumn{1}{r}{$.009$} & \multicolumn{1}{r}{$.006$} \\ 
&  & \multicolumn{1}{r}{} & \multicolumn{1}{r}{} & \multicolumn{1}{r}{} &  & 
\multicolumn{1}{r}{} & \multicolumn{1}{r}{} & \multicolumn{1}{r}{} \\ 
$\mu $ &  & \multicolumn{1}{r}{$.121$} & \multicolumn{1}{r}{$.102$} & 
\multicolumn{1}{r}{$.090$} &  & \multicolumn{1}{r}{$.096$} & 
\multicolumn{1}{r}{$.082$} & \multicolumn{1}{r}{$.072$} \\ 
$\nu $ &  & \multicolumn{1}{r}{$.124$} & \multicolumn{1}{r}{$.099$} & 
\multicolumn{1}{r}{$.087$} &  & \multicolumn{1}{r}{$.163$} & 
\multicolumn{1}{r}{$.144$} & \multicolumn{1}{r}{$.136$} \\ 
$\phi _{1}$ &  & \multicolumn{1}{r}{$.738$} & \multicolumn{1}{r}{$.515$} & 
\multicolumn{1}{r}{$.376$} &  & \multicolumn{1}{r}{$.436$} & 
\multicolumn{1}{r}{$.261$} & \multicolumn{1}{r}{$.188$} \\ 
$\phi _{2}$ &  & \multicolumn{1}{r}{$.882$} & \multicolumn{1}{r}{$.726$} & 
\multicolumn{1}{r}{$.558$} &  & \multicolumn{1}{r}{$.588$} & 
\multicolumn{1}{r}{$.389$} & \multicolumn{1}{r}{$.290$} \\ 
$\psi _{1}$ &  & \multicolumn{1}{r}{$.243$} & \multicolumn{1}{r}{$.249$} & 
\multicolumn{1}{r}{$.206$} &  & \multicolumn{1}{r}{$.138$} & 
\multicolumn{1}{r}{$.090$} & \multicolumn{1}{r}{$.061$} \\ 
$\psi _{2}$ &  & \multicolumn{1}{r}{$.216$} & \multicolumn{1}{r}{$.216$} & 
\multicolumn{1}{r}{$.176$} &  & \multicolumn{1}{r}{$.145$} & 
\multicolumn{1}{r}{$.097$} & \multicolumn{1}{r}{$.068$} \\ 
\multicolumn{1}{r}{} &  &  &  &  &  &  &  &  \\ 
$\sigma _{u1}$ &  & \multicolumn{1}{r}{$.065$} & \multicolumn{1}{r}{$.057$}
& \multicolumn{1}{r}{$.029$} &  & \multicolumn{1}{r}{$.039$} & 
\multicolumn{1}{r}{$.027$} & \multicolumn{1}{r}{$.020$} \\ 
$\sigma _{u2}$ &  & \multicolumn{1}{r}{$.065$} & \multicolumn{1}{r}{$.069$}
& \multicolumn{1}{r}{$.038$} &  & \multicolumn{1}{r}{$.033$} & 
\multicolumn{1}{r}{$.024$} & \multicolumn{1}{r}{$.018$} \\ 
$\sigma _{v1}$ &  & \multicolumn{1}{r}{$.070$} & \multicolumn{1}{r}{$.058$}
& \multicolumn{1}{r}{$.096$} &  & \multicolumn{1}{r}{$.062$} & 
\multicolumn{1}{r}{$.047$} & \multicolumn{1}{r}{$.036$} \\ 
$\sigma _{v2}$ &  & \multicolumn{1}{r}{$.071$} & \multicolumn{1}{r}{$.082$}
& \multicolumn{1}{r}{$.065$} &  & \multicolumn{1}{r}{$.037$} & 
\multicolumn{1}{r}{$.027$} & \multicolumn{1}{r}{$.018$} \\ 
$\sigma _{\eta }$ &  & \multicolumn{1}{r}{$.067$} & \multicolumn{1}{r}{$.081$%
} & \multicolumn{1}{r}{$.062$} &  & \multicolumn{1}{r}{$.012$} & 
\multicolumn{1}{r}{$.011$} & \multicolumn{1}{r}{$.010$} \\ 
&  & \multicolumn{1}{r}{} & \multicolumn{1}{r}{} & \multicolumn{1}{r}{} &  & 
\multicolumn{1}{r}{} & \multicolumn{1}{r}{} & \multicolumn{1}{r}{} \\ 
$u_{i1}$ &  & \multicolumn{1}{r}{$.217$} & \multicolumn{1}{r}{$.184$} & 
\multicolumn{1}{r}{$.170$} &  & \multicolumn{1}{r}{$.154$} & 
\multicolumn{1}{r}{$.140$} & \multicolumn{1}{r}{$.134$} \\ 
$u_{i2}$ &  & \multicolumn{1}{r}{$.163$} & \multicolumn{1}{r}{$.141$} & 
\multicolumn{1}{r}{$.121$} &  & \multicolumn{1}{r}{$.118$} & 
\multicolumn{1}{r}{$.104$} & \multicolumn{1}{r}{$.097$} \\ 
$v_{i1}$ &  & \multicolumn{1}{r}{$.167$} & \multicolumn{1}{r}{$.159$} & 
\multicolumn{1}{r}{$.162$} &  & \multicolumn{1}{r}{$.168$} & 
\multicolumn{1}{r}{$.151$} & \multicolumn{1}{r}{$.143$} \\ 
$v_{i2}$ &  & \multicolumn{1}{r}{$.153$} & \multicolumn{1}{r}{$.148$} & 
\multicolumn{1}{r}{$.138$} &  & \multicolumn{1}{r}{$.105$} & 
\multicolumn{1}{r}{$.083$} & \multicolumn{1}{r}{$.072$}%
\end{tabular}

\caption{Simulation Results. Root mean squared errors of estimators based on five-knot B-splines
under different baseline rates $r$ and sample sizes $n$, for model with variance proportion $\alpha = .75$.}%
\label{tab:Sim_errors}%
\end{table}%

As measures of estimation error we considered the root mean squared errors.
For scalar parameters, e.g.~$\sigma _{\eta }$, they are defined as usual: $%
E^{1/2}\{(\hat{\sigma}_{\eta }-\sigma _{\eta })^{2}\}$. For functional
parameters, e.g.~$\mu (x)$, they are defined in terms of the $L^{2}$-norm: $%
E^{1/2}(\Vert \hat{\mu}-\mu \Vert ^{2})\}^{1/2}$. For the random-effect
predictors, e.g.~the $\hat{u}_{i1}$s, they are defined as $%
E^{1/2}\{\sum_{i=1}^{n}(\hat{u}_{i1}-u_{i1})^{2}/n\}$. The sign of the $\hat{%
\phi}_{k}(x)$s and the $\hat{\psi}_{k}(x)$s, which in principle are
indeterminate, were chosen as the signs of the inner products $\langle \hat{%
\phi}_{k},\phi _{k}\rangle $ and $\langle \hat{\psi}_{k},\psi _{k}\rangle $;
the signs of the $\hat{u}_{ik}$s, $\hat{v}_{ik}$s and the elements of $%
\mathbf{\hat{\Sigma}}_{uv}$ were changed accordingly. For reasons of space
we only report here the results for the six sampling models with $\alpha
=.75 $, $n\leq 200$ and estimators obtained using five-knot splines (Table %
\ref{tab:Sim_errors}). The rest of the results can be found in the
Supplementary Material and are largely in line with the ones reported here.
Also given in the Supplementary Material are plots of the functional
estimators, which help assess the relative weight of the bias and variance
in the overall mean squared error.

We see in Table \ref{tab:Sim_errors} that the estimation errors decrease as $%
n$ increases, as expected, for both baseline rates $r$. However, the latter
has a big impact on the accuracy of the estimators, particularly of the
components $\phi _{1}$ and $\phi _{2}$. A look at the plots in the
Supplementary Material reveals that most of the error of $\hat{\phi}_{1}$
and $\hat{\phi}_{2}$ comes from the bias rather than the variance, and, for
a given $n$, the bias decreases fast as $r$ increases. Part of the bias of $%
\hat{\phi}_{1}$ and $\hat{\phi}_{2}$ can be attributed to component
reversal, which is more frequent for the models with $\alpha =.60$ than for $%
\alpha =.75$. This is also the case, but to a lesser degree, for $\hat{\psi}%
_{1}$ and $\hat{\psi}_{2}$, which, for each $(n,r)$ combination, are more
accurate estimators of their respective parameters than $\hat{\phi}_{1}$ and 
$\hat{\phi}_{2}$.

Table \ref{tab:Sim_AVs} compares the true finite-sample standard deviations
of the elements of $\mathbf{\hat{\Sigma}}_{uv}$ with their median asymptotic
approximations, and also provides median absolute errors of these
approximations, for the estimators based on five-knot splines and models
with variance proportion $\alpha =.75$; for $\alpha =.60$ and for ten-knot
splines the results are given in the Supplementary Material. The dimension
of $\mathbf{\theta }$ for five-knot splines is $s=63$, so Fisher's
information matrix estimator $\mathbf{\hat{F}}_{0}$ is singular for $n=50$;
thus we only report the results for $n\geq 100$. Overall, we see that the
asymptotic standard deviations are very accurate estimators of the true
standard deviations for $n\geq 200$. For ten-knot splines, where the
dimension of $\mathbf{\theta }$ is $s=93$, the tables in the Supplementary
Material show that the approximation is accurate for $n\geq 400$. This
suggests ratios $n/s\geq 3$ as sufficient for accurate asymptotic
approximations of the variances.

\begin{table}[tbp] \centering%

\begin{tabular}{ccccccccccccc}
&  & \multicolumn{11}{c}{$r=10$} \\ 
&  & \multicolumn{3}{c}{$n=100$} &  & \multicolumn{3}{c}{$n=200$} &  & 
\multicolumn{3}{c}{$n=400$} \\ 
Parameter &  & True & Med & MAE &  & True & Med & MAE &  & True & Med & MAE
\\ 
$\Sigma _{uv,11}$ &  & $.31$ & $.63$ & $.32$ &  & $.24$ & $.28$ & $.04$ &  & 
$.16$ & $.16$ & $.01$ \\ 
$\Sigma _{uv,21}$ &  & $.38$ & $.73$ & $.35$ &  & $.24$ & $.36$ & $.12$ &  & 
$.15$ & $.21$ & $.06$ \\ 
$\Sigma _{uv,12}$ &  & $.23$ & $.42$ & $.19$ &  & $.15$ & $.21$ & $.06$ &  & 
$.11$ & $.12$ & $.02$ \\ 
$\Sigma _{uv,22}$ &  & $.17$ & $.30$ & $.13$ &  & $.12$ & $.14$ & $.02$ &  & 
$.13$ & $.09$ & $.05$ \\ 
&  &  &  &  &  &  &  &  &  &  &  &  \\ 
&  & \multicolumn{11}{c}{$r=30$} \\ 
$\Sigma _{uv,11}$ &  & $.25$ & $.45$ & $.20$ &  & $.18$ & $.21$ & $.03$ &  & 
$.12$ & $.13$ & $.01$ \\ 
$\Sigma _{uv,21}$ &  & $.17$ & $.32$ & $.15$ &  & $.11$ & $.16$ & $.04$ &  & 
$.08$ & $.10$ & $.02$ \\ 
$\Sigma _{uv,12}$ &  & $.14$ & $.24$ & $.10$ &  & $.10$ & $.12$ & $.02$ &  & 
$.06$ & $.07$ & $.01$ \\ 
$\Sigma _{uv,22}$ &  & $.09$ & $.18$ & $.09$ &  & $.06$ & $.09$ & $.02$ &  & 
$.04$ & $.05$ & $.01$%
\end{tabular}

\caption{Simulation Results. True standard deviations and median and median absolute errors of
estimated asymptotic standard deviations ($\times 10$) of estimators
under different baseline rates $r$ and sample sizes $n$, for estimators based on five-knot B-splines 
and variance proportion $\alpha =.75$}\label{tab:Sim_AVs}%
\end{table}%

\section{\label{sec:Example}Application: online auction data}

The eBay auction data mentioned in Section \ref{sec:Introd} was downloaded
from the companion website of Jank and Shmueli (2010). In this sample there
were 194 items sold at auction, and each auction lasted seven days. A
subsample of 20 bid-price trajectories are shown in Figure \ref%
{fig:samp_data}. The dots are the actual bids; the solid lines were drawn
for better visualization. Figure \ref{fig:samp_data} shows that bidding
activity tends to concentrate at the beginning and at the end of the
auctions, in patterns that have been called `early bidding'\ and `bid
sniping', respectively. Some articles (e.g.~Backus et al., 2015) have
pointed out that `bid sniping'\ is annoying for bidders, and partly as a
consequence of this, the number of items auctioned at eBay has steadily
decreased over the years compared to the number of items sold at fixed
prices (Einav et al., 2015). It has been hypothesized that bid sniping is
triggered by the perception that an item's current bid price is low. We will
not establish causation here, since our models are not intended for that,
but the results obtained below are in line with this hypothesis.

\FRAME{ftbpFU}{3.5206in}{2.6498in}{0pt}{\Qcb{Online Auction Data. Price
trajectories of Palm Digital Assistants auctioned at eBay (first 20
trajectories in a sample of 194).}}{\Qlb{fig:samp_data}}{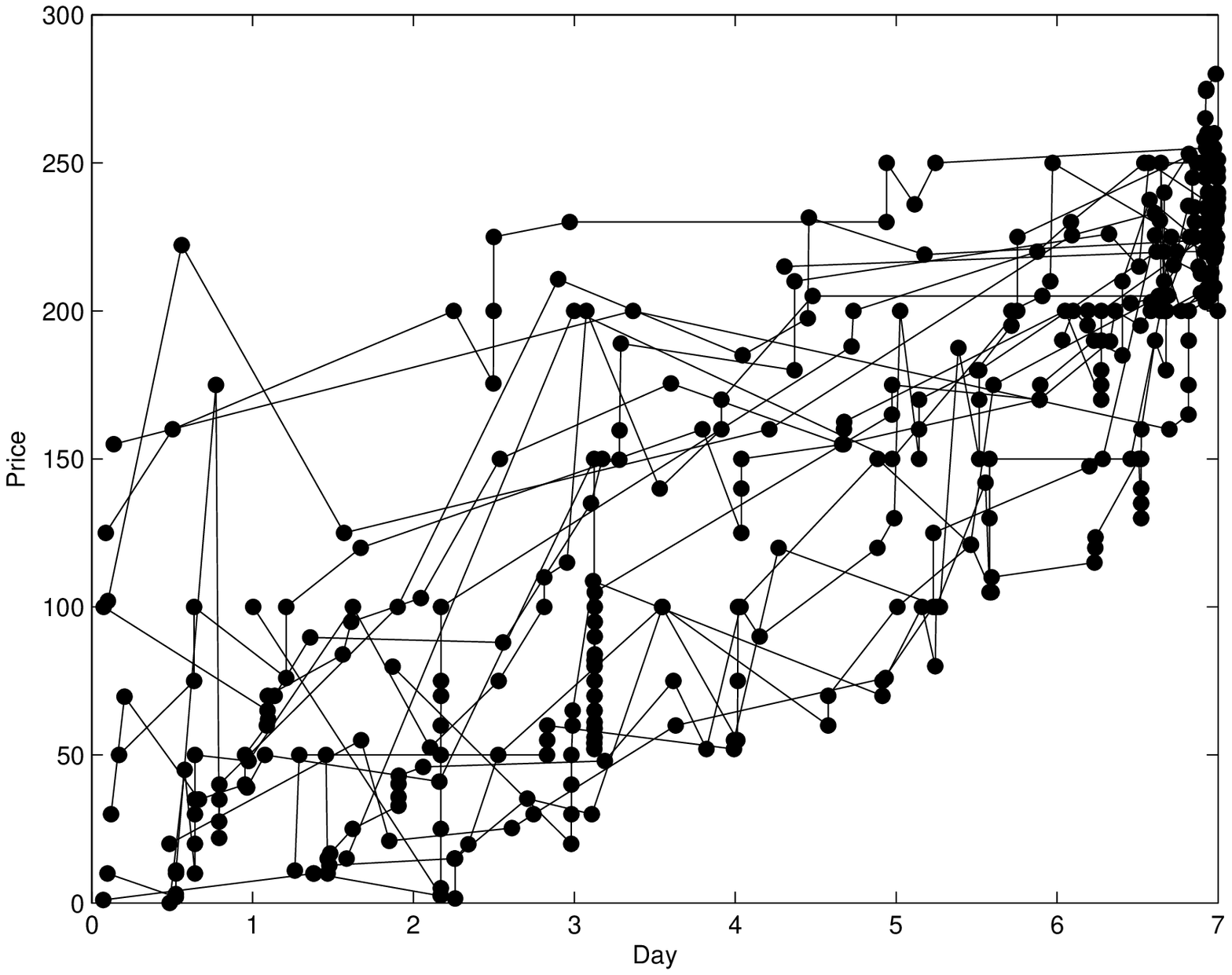}{%
\special{language "Scientific Word";type "GRAPHIC";maintain-aspect-ratio
TRUE;display "ICON";valid_file "F";width 3.5206in;height 2.6498in;depth
0pt;original-width 5.8219in;original-height 4.3708in;cropleft "0";croptop
"1";cropright "1";cropbottom "0";filename 'samp_data.eps';file-properties
"XNPEU";}}

To estimate the functional means and components we used cubic $B$-splines
with five equally spaced knots. We found the smoothing parameters
graphically (the plots can be found in the Supplementary Material),
obtaining $\xi _{1}=\xi _{2}=\xi _{4}=10^{-4}$ and $\xi _{3}=10^{-6}$. From
preliminary trial fits with five components for each process, we found that
the first two components of $X$ explain 77\% of the variability and the
first three components of $Y$ explain essentially 100\% of the variability
(the other two eigenvalues are negligible); therefore, we chose $p_{1}=2$
and $p_{2}=3$. The estimated mean and components are shown in Figure \ref%
{fig:mean_pcs}. Figure \ref{fig:mean_pcs}(a) shows the baseline intensity
function $\lambda _{0}(t)=\exp \mu (t)$ of the bidding process, and we see
that most of the bidding activity tends to occur towards the end of the
auction. Some items attract, overall, more bids than others, and this is
explained by the first component (Fig.~\ref{fig:mean_pcs}(c)): a positive
score on $\phi _{1}$ corresponds to an intensity function $\lambda $ above
the baseline. The second component is related to `bid sniping': for items
with positive scores on $\phi _{2}$, the number of bids in the last two days
of the auction will be above the mean. Regarding bid price, Fig.~\ref%
{fig:mean_pcs}(b) shows the mean price trajectory $\nu (t)$ and Fig.~\ref%
{fig:mean_pcs}(d) the components. The first component is associated with
price level: items with positive scores on $\psi _{1}$ will show prices
above the mean over the whole auction period. The second component is a
contrast: items with positive scores on $\psi _{2}$ tend to show prices
below the mean at the beginning of the auction and above the mean towards
the end.

\FRAME{ftbpFU}{5.2494in}{3.7291in}{0pt}{\Qcb{Online Auction Data. (a)
Baseline intensity function of bidding time process. (b) Mean price
trajectory. (c) Components of bidding time process, $\protect\phi _{1}$
(dashed line) and $\protect\phi _{2}$ (dash-dot line). (d) Components of
price trajectories, $\protect\psi _{1}$ (dashed line), $\protect\psi _{2}$
(dash-dot line) and $\protect\psi _{3}$ (dotted line).}}{\Qlb{fig:mean_pcs}}{%
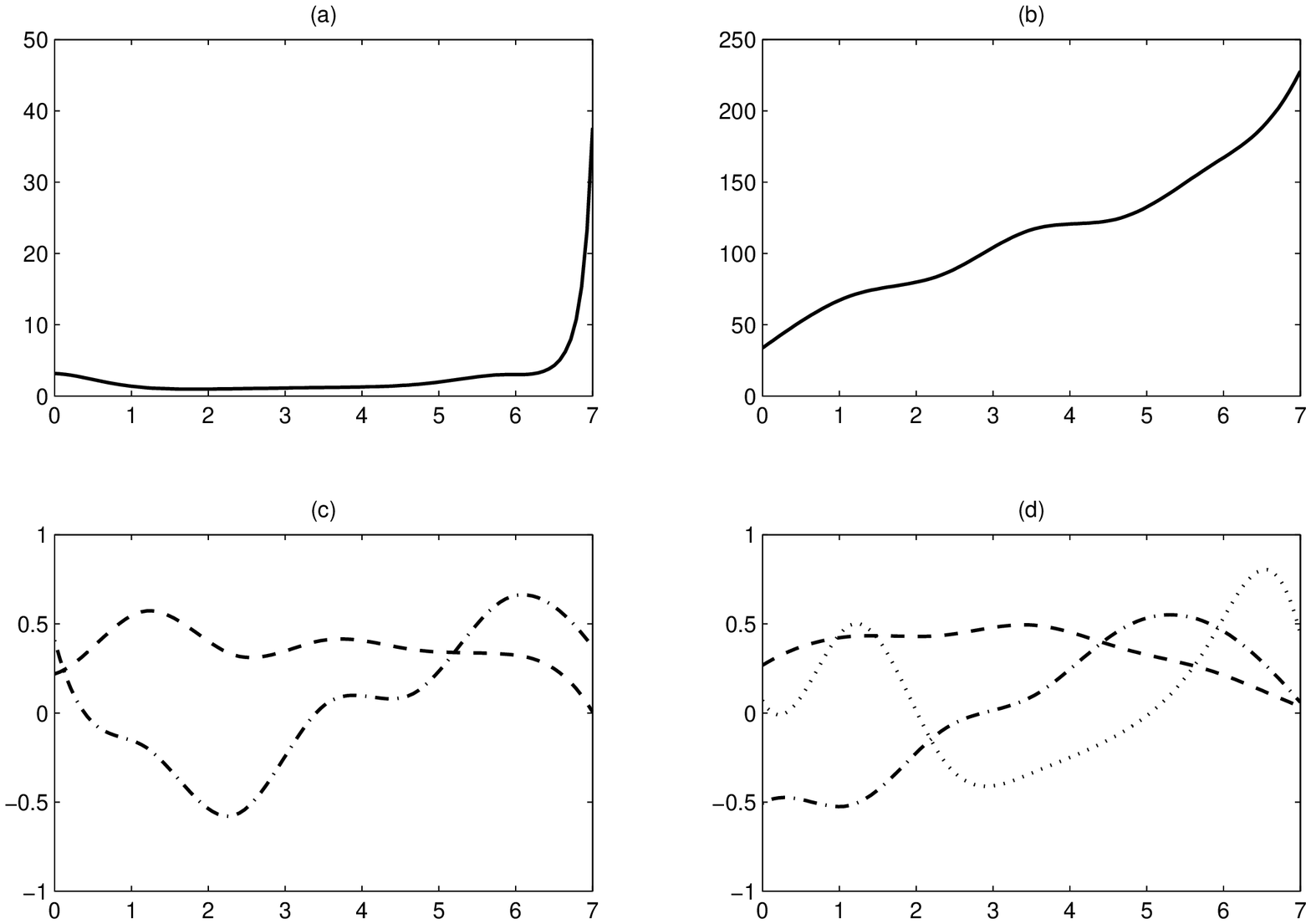}{\special{language "Scientific Word";type
"GRAPHIC";maintain-aspect-ratio TRUE;display "ICON";valid_file "F";width
5.2494in;height 3.7291in;depth 0pt;original-width 5.8219in;original-height
4.3708in;cropleft "0";croptop "1";cropright "1";cropbottom "0";filename
'mean_pcs.eps';file-properties "XNPEU";}}

The estimated cross-covariance and cross-correlation matrices were 
\[
\mathbf{\hat{\Sigma}}_{uv}=\left( 
\begin{array}{ccc}
-256.9 & 48.1 & 22.6 \\ 
-83.1 & -36.9 & -1.5%
\end{array}%
\right) \text{ and }\mathbf{\hat{\rho}}_{uv}=\left( 
\begin{array}{ccc}
-.69 & .41 & .28 \\ 
-.54 & -.77 & -.05%
\end{array}%
\right) . 
\]%
The asymptotic standard deviations of the elements of $\mathbf{\hat{\Sigma}}%
_{uv}$ obtained from Theorem \ref{thm:Asymp} and bootstrap standard
deviations based on $100$ wild bootstrap replications were 
\[
\mathrm{sd}_{\mathrm{asymp}}(\mathbf{\hat{\Sigma}}_{uv})=\left( 
\begin{array}{ccc}
73.3 & 17.7 & 9.9 \\ 
20.5 & 6.8 & 5.7%
\end{array}%
\right) \text{ and }\mathrm{sd}_{\mathrm{boot}}(\mathbf{\hat{\Sigma}}%
_{uv})=\left( 
\begin{array}{ccc}
76.7 & 18.3 & 13.4 \\ 
22.3 & 7.5 & 5.3%
\end{array}%
\right) , 
\]%
which are very similar to one another. We can conclude that all correlations
involving the first two components of each process are statistically
significant but none of the correlations involving $\psi _{3}$ are.

Figure \ref{fig:pc_scores} shows scatter plots of the estimated random
effects $\hat{u}_{ik}$s versus $\hat{v}_{ik}$s for the significant
components. Normal probability plots of the component scores and the
residuals $\hat{\eta}_{ij}$s are shown in the Supplementary Material. The
component scores appear to be largely Gaussian; only the $\hat{u}_{i1}$s
show a mild departure from normality. The residuals $\hat{\eta}_{ij}$s show
tails somewhat heavier than Normal, but no gross outliers are evident.

These results are in line with intuition. The negative correlations between $%
v_{1}$ and both $u_{1}$ and $u_{2}$ show that items with perceived low
prices tend to attract more bidders and trigger bid sniping. The strong
negative correlation between $u_{2}$ and $v_{2}$ shows that bid snipping is
particularly associated with price trajectories that are found to be well
below the mean on the fifth day of the auction.

\FRAME{ftbpFU}{5.5633in}{3.9505in}{0pt}{\Qcb{Online Auction Data.
Scatterplots of component scores of the bidding time process versus
component scores of price trajectories.}}{\Qlb{fig:pc_scores}}{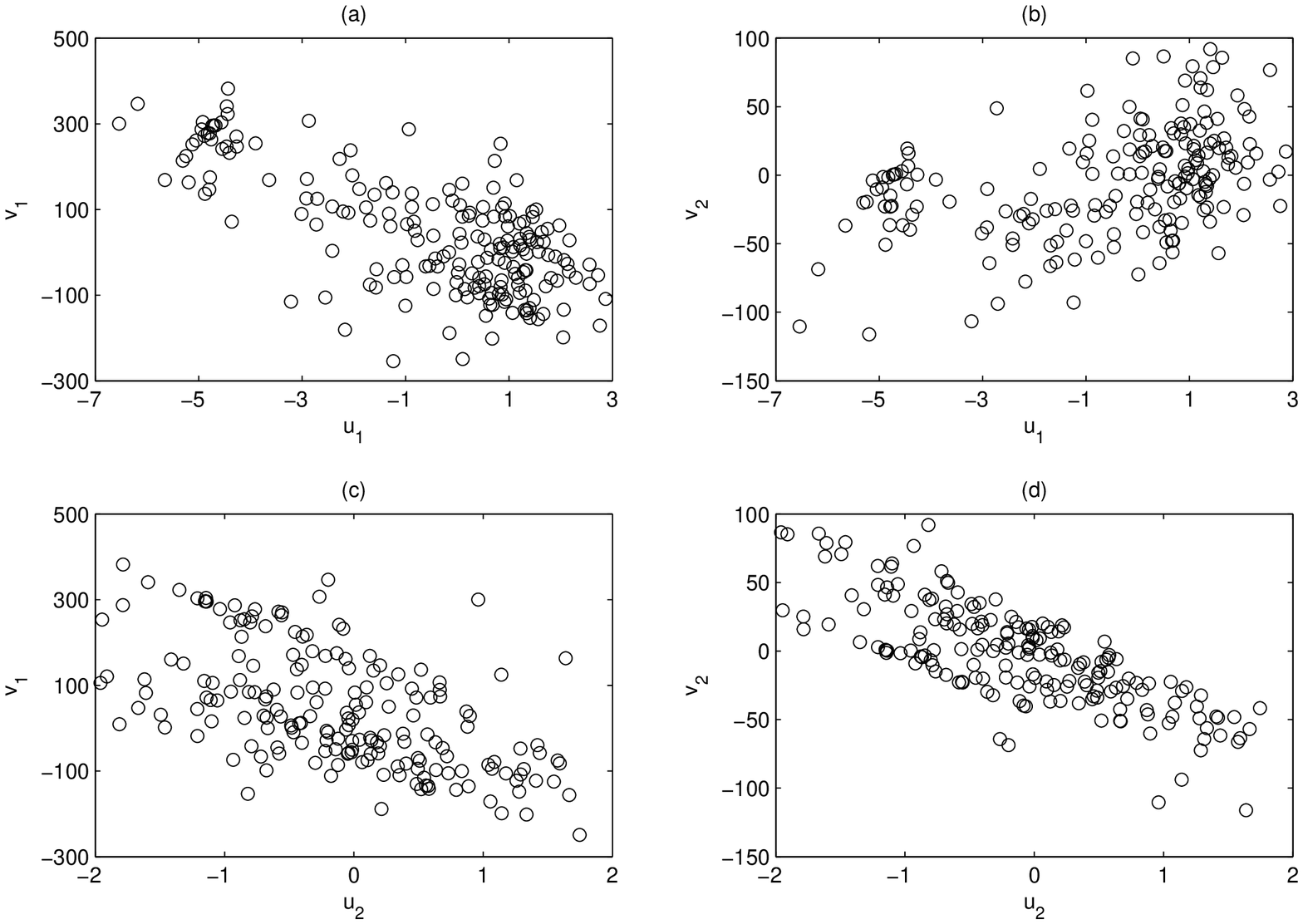%
}{\special{language "Scientific Word";type "GRAPHIC";maintain-aspect-ratio
TRUE;display "ICON";valid_file "F";width 5.5633in;height 3.9505in;depth
0pt;original-width 5.8219in;original-height 4.3708in;cropleft "0";croptop
"1";cropright "1";cropbottom "0";filename 'pc_scores.eps';file-properties
"XNPEU";}}

To illustrate with a few specific cases, Figure \ref{fig:pc_minmax} shows
the price trajectories of items with largest and smallest scores $v_{1}$ and 
$v_{2}$. Figure \ref{fig:pc_minmax}(a) shows the item with largest $v_{1}$
score, and consequently low $u_{1}$ score: an expensive item that attracted
only two bets. Figure \ref{fig:pc_minmax}(b) shows the opposite, the item
with lowest $v_{1}$ score and consequently large $u_{1}$ and $u_{2}$ scores:
an underpriced item that attracted a lot of bids towards the end of the
auction, a typical case of bid sniping. Figure \ref{fig:pc_minmax}(c) shows
the item with largest $v_{2}$ score, and consequently large $u_{1}$ score
but low $u_{2}$ score: and item that started off with a low price and
attracted many bids at the beginning of the auction, which sent the price
above the mean early in the auction period and then did not attract many
late bidders. Figure \ref{fig:pc_minmax}(d), the item with lowest $v_{2}$
score, shows the opposite situation: the few bids placed at the beginning of
the auction period were well above the mean, but towards the end some lower
bids are placed (an unusual but possible situation) which triggered bid
snipping.

\FRAME{ftbpFU}{5.5002in}{4.0655in}{0pt}{\Qcb{Online Auction Data. Estimated
price trajectories (solid line) and mean price trajectory (dashed line)
along with actual bets (asterisks) for items with (a) largest score on first 
$Y$-component, (b) lowest score on first $Y$-component, (c) largest score on
second $Y$-component, and (d) lowest score on second $Y$-component.}}{\Qlb{%
fig:pc_minmax}}{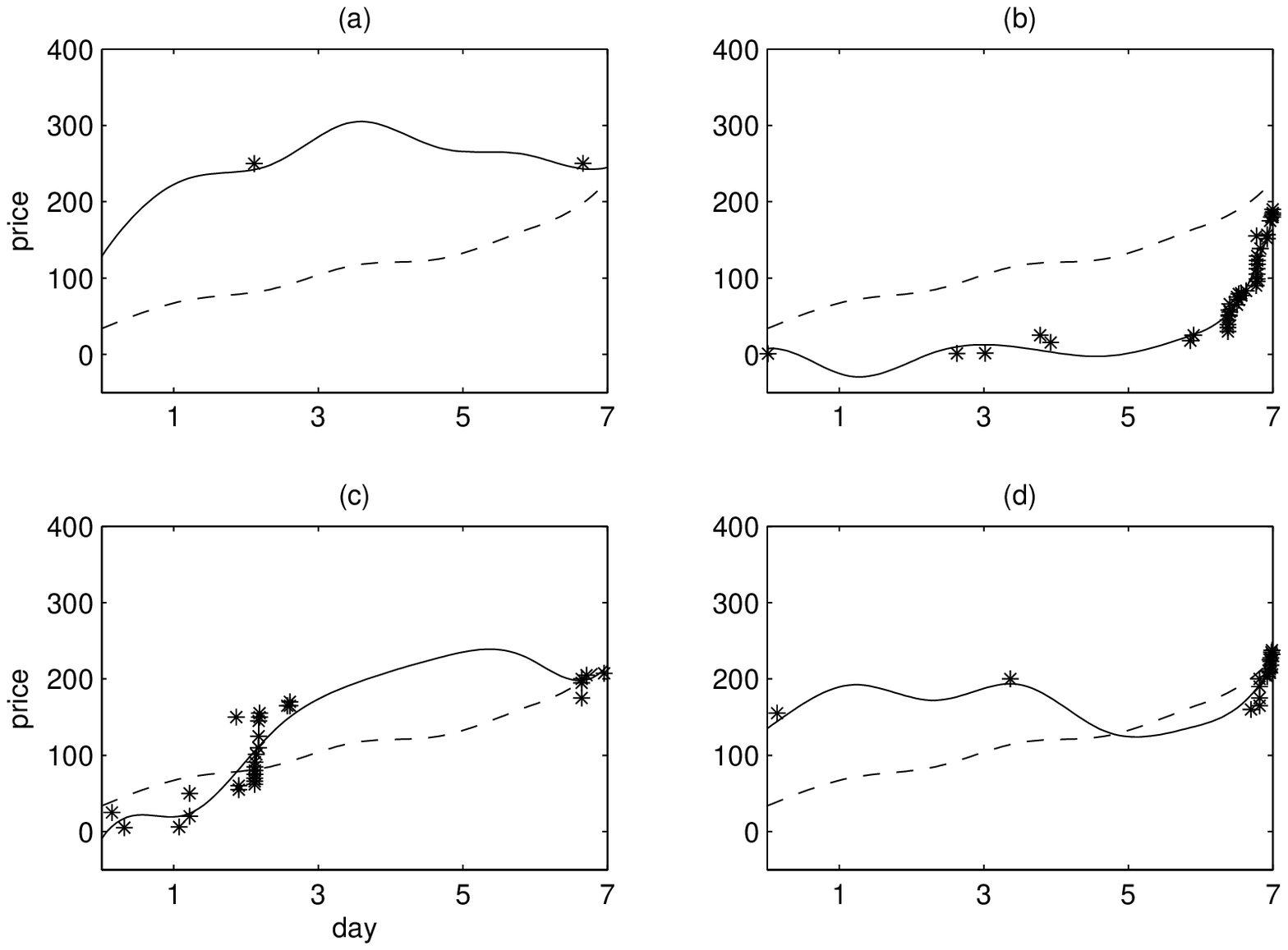}{\special{language "Scientific Word";type
"GRAPHIC";maintain-aspect-ratio TRUE;display "ICON";valid_file "F";width
5.5002in;height 4.0655in;depth 0pt;original-width 6.4714in;original-height
4.7824in;cropleft "0";croptop "1";cropright "1";cropbottom "0";filename
'pc_minmax.eps';file-properties "XNPEU";}}

\section{Discussion}

In this paper we have presented a unified model for the joint statistical
analysis of a functional response variable and the distribution of the grid
points at which the variable is measured. Although the problems of
estimating sparse functional data and intensity functions of point processes
had been considered in the literature, that had been done separately up to
this point. Work on canonical correlation analysis for sparse data (Shin and
Lee, 2015) is not really applicable in this setting, because instead of two
sparsely observed functional variables we have a single functional variable
and a random grid, which involves a completely different model and
estimation process.

Our model allows statistical inference for the correlations between
components of the grid-point process and the response variable. For this we
have developed a parametric asymptotic theory in Section \ref%
{sec:Asymptotics}, where $\sqrt{n}$-consistency is obtained but at the price
of ignoring asymptotic bias. When the latter is negligible, for example when
the target functions are smooth and the basis family used for estimation is
large enough, the asymptotic approximation is very accurate, as we showed by
simulation and example in Sections \ref{sec:Simulations} and \ref%
{sec:Example}. However, if the target functions were more irregular and the
asymptotic bias more significant, a truly nonparametric asymptotics with the
dimension of the basis family growing with $n$ would be more appropriate,
although the rate of convergence would be lower than $\sqrt{n}$. This is
still an open problem.

The model in Section \ref{sec:Model} uses latent variables whose
distributions are assumed Normal. Of course this is always going to be an
approximation at best. While mild departures from normality may not affect
the validity of the results, more serious deviations like gross outliers or
very heavy-tailed distributions most likely will. For reasons of space we
could not embark on a thorough robustness analysis in this paper, but the
model and maximum likelihood estimators we proposed can be easily modified
to accommodate heavier-tailed distributions, like Student's $t$
distributions, for the latent variables. This is also a matter for future
research.

\section{Acknowledgement}

This research was partly supported by US National Science Foundation grant
DMS 1505780.

\section*{References}

\begin{description}
\item Arribas-Gil, A., and M\"{u}ller, H.-G. (2014). Pairwise dynamic time
warping for event data. \emph{Computational Statistics and Data Analysis }%
\textbf{69} 255--268.

\item Ash, R.B. and Gardner, M.F. (1975). \emph{Topics in stochastic
processes}. Academic Press, New York.

\item Backus, M., Blake, T., Masterov, D.V., and Tadelis, S. (2015). Is
sniping a problem for online auction markets? \emph{NBER Working Paper}
No.~20942.

\item Baddeley, A. (2007). Spatial point processes and their applications.
In \emph{Stochastic Geometry}, Lecture Notes in Mathematics 1892, pp.~1--75.
Springer, New York.

\item Baddeley, A. (2010). Multivariate and marked point processes. In A. E.
Gelfand, P. J. Diggle, P. Guttorp and M. Fuentes (eds), \emph{Handbook of
Spatial Statistics}, CRC Press, Boca Raton, pp. 299--337.

\item Barrett, J., Diggle, P., Henderson, R. and Taylor-Robinson, D. (2015).
Joint modelling of repeated measurements and time-to-event outcomes:
flexible model specification and exact likelihood inference. \emph{Journal
of the Royal Statistical Society: Series B} \textbf{77 }131--148.

\item Cox, D.R., and Isham, V. (1980). \emph{Point Processes.} Chapman and
Hall/CRC, Boca Raton.

\item Dempster, A.P., Laird, N.M., and Rubin, D.B. (1977). Maximum
likelihood from incomplete data via the EM algorithm. \emph{Journal of the
Royal Statistical Society Series B} \textbf{39} 1--38.

\item Einav, L., Farronato, C., Levin, J.D., and Sundaresan, N. (2015).
Sales mechanisms in online markets: What happened to Internet auctions? 
\emph{NBER Working Paper} No.~19021.

\item Gervini, D. (2016). Independent component models for replicated point
processes. \emph{Spatial Statistics }\textbf{18} 474--488.

\item Geyer, C.J. (1994). On the asymptotics of constrained M-estimation. 
\emph{The Annals of Statistics }\textbf{22} 1993--2010.

\item Guan, Y., and Afshartous, D. R. (2007). Test for independence between
marks and points of marked point processes: a subsampling approach. \emph{%
Environmental and Ecological Statistics }\textbf{14} 101--111.

\item James, G., Hastie, T. G. and Sugar, C. A. (2000). Principal component
models for sparse functional data.\ \emph{Biometrika} \textbf{87} 587--602.

\item Jank, W., and Shmueli, G. (2006). Functional data analysis in
electronic commerce research. \emph{Statistical Science }\textbf{21}
155--166.

\item Jank, W., and Shmueli, G. (2010). \emph{Modeling Online Auctions.}
Wiley \& Sons, New York.

\item Knight, K., and Fu, W. (2000). Asymptotics for lasso-type estimators. 
\emph{The Annals of Statistics }\textbf{28} 1356--1378.

\item M\o ller, J., and Waagepetersen, R.P. (2004). \emph{Statistical
Inference and Simulation for Spatial Point Processes}. Chapman and Hall/CRC,
Boca Raton.

\item M\o ller, J., Ghorbani, M., and Rubak, E. (2016). Mechanistic
spatio-temporal point process models for marked point processes, with a view
to forest stand data. \emph{Biometrics }\textbf{72} 687--696.

\item M\"{u}ller, H.G. (2008). Functional modeling of longitudinal data. 
\emph{Longitudinal data analysis} \textbf{1} 223--252.

\item Myllym\"{a}ki, M., Mrkvi\v{c}ka, T., Seijo, H. and Grabarnik, P.
(2017). Global envelope tests for spatial processes. \emph{Journal of the
Royal Statistical Society Series B} \textbf{79} 381--404.

\item Pollard, D. (1984). \emph{Convergence of Stochastic Processes. }%
Springer, New York.

\item Ramsay, J.O., and Silverman, B.W. (2005). \emph{Functional Data
Analysis (second edition)}. Springer, New York.

\item Rathbun, S. L. and Shiffman, S. (2016). Mixed effects models for
recurrent events data with partially observed time-varying covariates:
Ecological momentary assessment of smoking. \emph{Biometrics }\textbf{72 }%
46--55.

\item Rice, J.A. (2004). Functional and longitudinal data analysis:
perspectives on smoothing. \emph{Statistica Sinica }\textbf{14} 631--647.

\item Rockafellar, R.T., and Wets, R.J. (1998). \emph{Variational Analysis.}
Springer, New York.

\item Ruppert, D. (2002). Selecting the number of knots for penalized
splines. \emph{Journal of Computational and Graphical Statistics} \textbf{11}
735--757.

\item Scheike, T.H. (1997). A general framework for longitudinal data
through marked point processes. \emph{Biometrical Journal }\textbf{39}
57--67.

\item Shin, H., and Lee, S. (2015). Canonical correlation analysis for
irregularly and sparsely observed functional data. \emph{Journal of
Multivariate Analysis} \textbf{134} 1--18.

\item Shmueli, G., and Jank, W. (2005). Visualizing online auctions. \emph{%
Journal of Computational and Graphical Statistics} \textbf{14} 299--319.

\item Streit, R.L. (2010). \emph{Poisson Point Processes: Imaging, Tracking,
and Sensing.} Springer, New York.

\item Van der Vaart, A. (2000). \emph{Asymptotic Statistics}. Cambridge
University Press, Cambridge, UK.

\item Wu, S., M\"{u}ller, H.-G., and Zhang, Z. (2013). Functional data
analysis for point processes with rare events. \emph{Statistica Sinica }%
\textbf{23} 1--23.

\item Xun, X., Cao, J., Mallick, B., Maity, A., and Carroll, R.J. (2013).
Parameter estimation of partial differential equations. \emph{Journal of the
American Statistical Association} \textbf{108} 1009--1020.

\item Yao, F., M\"{u}ller, H.-G. and Wang, J.-L. (2005). Functional linear
regression analysis for longitudinal data. \emph{The Annals of Statistics }%
\textbf{33 }2873--2903.

\item Yu, Y., and Ruppert, D. (2002). Penalized spline estimation for
partially linear single-index models. \emph{Journal of the American
Statistical Association} \textbf{97} 1042--1054.
\end{description}

\end{document}